\documentclass[twocolumn,aps,10pt,showpacs,showkeys,]{revtex4}
\usepackage[dvips]{graphicx}
\usepackage{amssymb}
\begin{document}
\title{Specific heat and orbital moment of CoO from first-principles atomistic calculations}
\author{R. J. Radwanski}
\affiliation{Center of Solid State Physics, S$^{nt}$Filip 5,
31-150 Krakow, Poland\\
Institute of Physics, Pedagogical University, 30-084 Krakow, Poland}
\author{Z. Ropka}
\affiliation{Center of Solid State Physics, S$^{nt}$Filip 5,
31-150 Krakow, Poland} \homepage{http://www.css-physics.pl}
\email{sfradwan@cyf-kr.edu.pl}

\begin{abstract}
We have for the first time calculated low-energy electronic structure both in paramagnetic and magnetic state as well as zero-temperature properties and thermodynamics. We consistently described magnetic properties of CoO in agreement with its insulating ground state. The orbital moment of 1.42 $\mu_B$ gives 35\% contribution to the total moment of 4.04 $\mu_B$ at T =0 K. We have calculated from this low-energy electronic structure the temperature dependence of the specific heat being in nice agreement with experimental data. In our approach CoO is an insulator independently on distortions and the magnetic order.

\end{abstract}
\maketitle \vspace {-0.3 cm}
\section{Introduction}
Understanding CoO and other 3$d$ monooxides is important for the
development of the solid-state theory. Their magnetism and the
insulating ground state is a subject of a fundamental controversy
by more than 70 years. Most models basing on the conventional band
theory predict FeO, CoO and NiO, having incomplete 3$d$ shell, to
be a metal, whereas they are observed to be very good insulators,
if stoichiometric.

Despite of 70 years of intense studies the problem
of the magnetism and the insulating ground-state of open-shell
transition metal oxides is still under debate \cite{1,2,3,4,5,6,7,8,9,10,11,12,13}.
We have largely solved this problem working in the purely ionic picture (we call this approach Quantum Atomistic Solid State Theory QUASST). In this paper we present theoretical description of the magnetism and low-energy electronic structure of CoO and the temperature dependence of the specific heat resulting from this structure. We earlier have described 
NiO \cite{14,15,16} and FeO \cite{17} followed by earlier
description of FeBr$_{2}$ \cite{18} and LaCoO$_{3}$ \cite{19}. In Refs 14-16 we
have described NiO within the localized atomistic paradigm with
only three parameters: the octupolar crystal-field parameter
$B_{4}$ =+21 K (10Dq =1.086 eV), the spin-orbit coupling
$\lambda_{s-o}$ = -480 K and a small trigonal distortion
$B_{2}^{0}$ =+50 K \cite{14,15,16}. These parameters are related to the lowest 21-fold degenerated $^3F$ term.  
\begin{figure}[t]
\vspace {0.0 cm}
\begin{flushleft}
\includegraphics[width =8.35 cm]{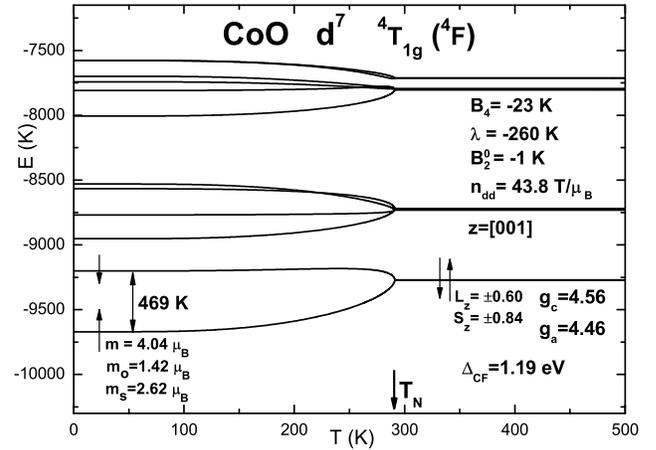}
\end{flushleft}
\vspace {-0.4 cm} \caption{Calculated splitting of the twelve 
lowest states of the $^{4}T_{1g}$ subterm for the Co$^{2+}$ ion (3$d^{7}$ configuration) both in the paramagnetic and magnetic state below T$_{N}$ of 291 K.  The used parameters are specified in the figure. The moment of the ground state of 4.04 $\mu_{B}$ is built up from the spin moment of 2.62 $\mu_{B}$ and
the orbital moment of 1.42 $\mu_{B}$; the splitting in the
paramagnetic state of the highest states is due to the spin-orbit 
coupling - an effect of the tetragonal distortion is very small causing only 
the thicking of the second and third levels. \vspace {-0.4 cm}}
\end{figure}
A fourth parameter, the molecular-field
coefficient $n_{dd}$ of $-200~ T/\mu_{B}$ describes the magnetic state. This $n_{dd}$ value yields $T_{N}$ of 525 K in good agreement with the experimental
observation for NiO. 

A direct motivation for this paper is presentation by Kant {\it et al.} \cite{20} of new measurements of the specific heat of CoO with an attempt to describe properties of CoO within an ionic model. Also works of Haverkort {\it et al.} \cite{21} and Wdowik and Parlinski \cite{22} from the last year contribute to studies of CoO.  
\section{Theoretical outline and results}
We have performed detailed calculations of the low-energy electronic structure related to the 28-fold degenerated $^4F$ term \cite{23} both in the paramagnetic and magnetic state. We took into account the experimental crystallographic structure - the $NaCl$-type structure with a=424 pm and a tetragonal distortion with c/a=0.988. 
We have calculated the strenght of the octahedral crystal field 10$Dq$=1.19 eV ($B_4$= -23 K = -2.0 meV - the minus sign is related with the octahedral position of the Co$^{2+}$ ion in the O$^{2-}$ octahedron; the calculated octupolar charge moment of surrounding charges at the Co site $A_{4}$ = +267 Ka$_{B}^{-4}$, a$_{B}$ is the Bohr radius; $\langle{r_{d}^{4}}\rangle$ for the Co$^{2+}$ ion is 12.6 a$_{B}^{4}$).

Calculations have been performed within the same scheme as for FeBr$_2$ \cite{18} and for NiO \cite{14,15,16} and other our earlier calculations for 4f/5f  compounds \cite{24}. In Hamiltonian we take into account the crystal-field interactions (the octahedral crystal field with a small tetragonal distortion $B_2^0$), the spin-orbit interactions $\lambda_{s-o}$ and inter-site spin-dependent interactions ($n_{dd}$). The resulting low-energy electronic structure is shown in Fig. 1. In
the magnetically-ordered state there appears a spin-gap of 466 K
(=40 meV). The ordered magnetic moment amounts at T= 0 K to 4.04 $\mu_{B}$ - it is built from the spin
moment of 2.62 $\mu_{B}$ and the orbital moment of 1.42
$\mu_{B}$ \cite{25}. The calculated orbital moment gives 35\% contribution to the total moment. The experimentally derived magnetic moment
amounts to 4.0 $\mu_{B}$ \cite{26}.

\begin{figure}
\begin{center}
\includegraphics[width =8.6 cm]{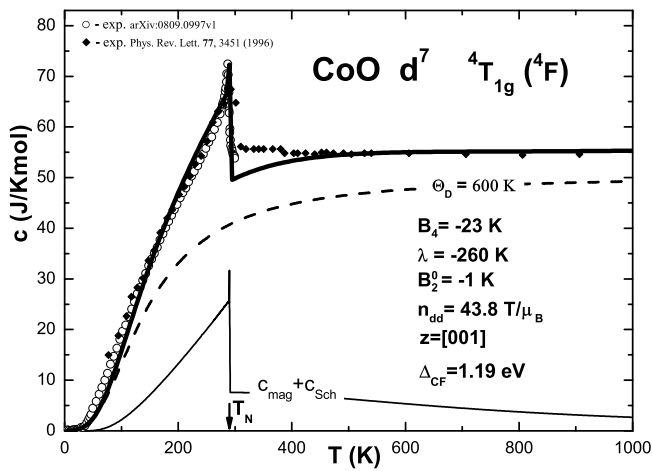}
\end{center} \vspace {-0.6 cm}
\caption{The calculated temperature dependence of the specific heat of CoO (thick line) contains the magnetic/Schottky contribution, the lattice and dilation contributions; thin line - the 3$d$ contribution $c_{d}(T)$ containing both magnetic and Schottky contributions. The dashed line shows the lattice contribution with single $\theta_D$ of 600 K. This lattice contribution approaches a value of 6 R, R is a gas constant, for
high temperatures according to the Dulong-Petit law. The dilation contribution is approximated by a linear term with temperature with $\gamma_d$ =3.4 mJ/K$^2$ mol. Points are experimental data: o - taken from Ref. \cite {20}, $\blacklozenge$ - taken from Ref. \cite{27}.}.\vspace {-0.5cm}
\end{figure}

The tetragonal distortion is small and causes a
slight splitting of the excited quartets. But this small
distortion determines in our calculations the direction of the Co
magnetic moment, in the CoO case with the c-axis compression, along the tetragonal direction. 

The formation of the magnetic state we have described for
numerous compounds - let mention exemplary 4f/3d/5f compounds
ErNi$_{5}$, FeBr$_{2}$, NiO, FeO, UPd$_{2}$Al$_{3}$ and
UGa$_{2}$, results of which have been published starting from
1992. In all these cases the magnetic energy is much smaller than
the overall CEF splitting. They are both ionic (FeBr$_{2}$
\cite{18}, NiO, FeO \cite{11}) and intermetallic (ErNi$_{5}$,
UPd$_{2}$Al$_{3}$, UGa$_{2}$) \cite{24} compounds. A derived
value of n= -43.8 T/$\mu_{B}$ (= -29.4 K) in CoO yields the molecular
field acting on the Co-ion moment at T = 0 K as 176 T - it
corresponds to the exchange field acting on the Co spin moment of
271 T and to the exchange energy of 41.2 meV. 

The calculated electronic structure very well describes the temperature
dependence of the specific heat as is shown in Fig. 2, where the calculated data are compared with experimental data. 
The calculated temperature dependence of the specific heat of CoO contains three contributions: the 3$d$ contribution $c_{d}(T)$ containing both magnetic and Schottky contributions, the lattice and dilation contributions. The magnetic/Schottky contribution is calculated from the derived electronic structure shown in Fig. 1 similarly to that discussed in Ref. \cite{18} for FeBr$_2$. The dilation contribution is linear with temperature \cite{27} - from thermal expansion and the bulk modulus we calculated its temperature coefficient as 3.4 mJ/K$^2$ mol (we denote it as $\gamma_d$ to be distinguished from the electronic Sommerfeld coefficient).
The lattice contribution we approximate by the Debye function. We did not have intention to fit perfectly the data but using a value of the Debye temperature $\theta_{D}$ of 600 K the experimental data are nicely reproduced in the very wide temperature range, from 0 to 1000 K. Here we give some values. At T=200 K $c_m$ =13.27, $c_{lat}$ =33.09 and $c_{dil}$ =0.68 J/Kmol gives total heat of 47.02 J/Kmol, which should be compared with 46.03 \cite{27} (46.63 \cite{20}) J/K mol.   
At T=400 K $c_m$ =$c_{Sch}$ =7.15, $c_{lat}$ =44.78 and $c_{dil}$ =1.36 J/Kmol gives total heat of 53.29 J/Kmol, compared to 54.80 J/Kmol \cite{27}. 
At T=600 K $c_m$ (=$c_{Sch}$) =5.35, $c_{lat}$ =47.62 and $c_{dil}$ =2.04 J/Kmol gives total heat of 55.01 J/Kmol, compared to 54.80 J/Kmol \cite{27}. All of these values are very close to experimental data of Refs \cite{20,27} as seen in Fig. 2.

The derived Debye temperature $\theta_{D}$ of 600 K is 
quite similar to a value for $\theta_{D}$ of 650 K we have obtained for NiO \cite{15} - it is a very plausible result owing to the structural and
compositional similarity of both monoxides.  

We would like to note that all of the used by us parameters
(dominant octahedral CEF parameter 10$Dq$ of 1.19 eV (or $B_{4}$= -23 K), the spin-orbit coupling $\lambda_{s-o}$= -260 K, lattice distortions B$_{2}^{0}$
= -1 K) have clear physical meaning. The most important
assumption is the existence of very strong correlations among
3$d$ electrons what assure the preservation of the atomistic
ionic integrity of the Co$^{2+}$ ion in the solid CoO. We stress
the good reproduction of experimental results like temperature
dependence of the heat capacity with a $\lambda$-type peak, the
value of the magnetic moment and, in particular, its direction.

The crystal-field interactions are relatively strong in CoO but
not so strong to destroy the ionic integrity of the 3$d$
electrons - it forms conditions for adequancy of the QUASST theory. The calculated value 10$Dq$ of 1.19 eV places CoO at value of 1.0 at the Tanabe-Sugano diagram drawn with $Dq$/B as abscissa \cite {28,29}. 

\section{Conclusions}
We have for the first time calculated low-energy electronic structure both in paramagnetic and magnetic state as well as zero-temperature properties and thermodynamics. We consistently described magnetic properties of CoO in agreement with its insulating ground state. The orbital moment of 1.42 $\mu_B$ gives 35\% contribution to the total moment of 4.04 $\mu_B$ at T =0 K. We have calculated from this low-energy electronic structure the temperature dependence of the specific heat being in nice agreement with experimental data. In our approach CoO is an insulator independently on distortions and the magnetic order.


\begin{thebibliography}{9}
\bibitem{23} In Refs \cite{20} and \cite{21} calculations have been performed only for 12 lowest states.
\end{thebibliography}
\end{document}